# Novel Molecular Signaling Method and System for Molecular Communication in Human Body

AbdulAziz Al-Helali1, Ben Liang, *Fellow, IEEE*, and Nidal Nasser, *Senior Member, IEEE*

*Abstract*— For millions of years, MC has been the primary method of communication in living organisms, cells, and the human body. Recently, Molecular Communication (MC) has been recognized as an enabling technology for nanonetworks [1] where MC is envisioned to enable nanorobots to achieve sophisticated and complex tasks in the human body for promising medical applications. Many MC methods that can be applied in the human body have been proposed and modeled in the literature. However, none of them can be used to convey information between distant points that are separated by body fluids, tissues, or placed in different organs. In this paper, we propose a new method and system for Molecular Communication in the Body, denoted MoSiMe and MoCoBo, respectively. The method takes advantage of how the absorption, distribution, metabolization, and elimination (ADME) bodily processes affects drugs, referred to as Pharmacokinetics (PK) in pharmacology, to enable MC between any points of the human body regardless of their distance even if they are in different parts of the body and separated by tissues and fluids. The architecture, design and components of the MoCoBo system and MoSiMe method are described and different transmitter designs, including a novel passive transmitter design for the first time in telecommunications, are introduced. An analytical model for the body channel is derived and validated with respect to existing human and animal tests. The model captures the ADME bodily processes that affect the kinetics of substances administered to the body. Additionally, an experimental MoCoBo proof of concept platform, capable of reliably sending and receiving a stream of bits between a transmitter and a receiver, is built and validated against clinical trials, animal tests and analytical models. The introduced platform can also be utilized to test modulation techniques and designs for new MoCoBo transmitters and receivers.

## I. Introduction

In recent years, there is an increased interest in Molecular Communication (MC) since it has been recognized as an enabling technology for nanonetworks [1]. Unlike classical forms of communications, such as electro-magnetic communication, MC uses molecules to convey information between the points of communication link, which for example could be a swarm of robots or larger scale devices that uses MC as an alternative link when classical forms of communications are not suitable.

For millions of years, MC has been the primary method of communication in living organisms, cells, and the human body. Cell signaling [2] and quorum sensing [3] are two examples of MC. In cell signaling, cells exchange molecules between each other to achieve specific goals such as control of growth and cell functionalities. Similarly, bacteria use MC in a process called quorum sensing to regulate expression of genes in response to the concentration of specific molecules called autoinducers which are used for managing many physiological activities such as interdependence, motility, and virulence [3]. Quorum sensing enables bacteria to communicate within a community and across species which give them the abilities of more developed organisms by taking coordinating actions [3].

Like quorum sensing in bacteria, by forming body area nanonetworks [4], MC is envisioned to enable nanorobots to achieve sophisticated and complex tasks in the human body for promising medical purposes such as disease detection, and targeted drug delivery [5]. For example, in disease detection, bacteriobots [6], which are made of modified flagellated bacteria, are used in targeting and treating tumor cells. The effectiveness of such bacteriobots can be greatly improved to perform tasks such as tumor early detection if they are interconnected [7] in a mobile ad hoc molecular nanonetwork (MAMNET) [8] to exchange information.

Many MC methods between nanorobots in the human body have been proposed and modeled in the literature. These methods can be broadly classified, based on distance, into short (nm-µm), medium (µm-mm), and long range (mm-m) [9]. Existing short and medium range methods such as free diffusion [10] and bacteria assisted propagation [11] can be used for communication in the scale of a few millimeters or less. Therefore, they limit the range of communication in one place such as inside a cell, tissue, or an organ. On the other hand, long range MC methods such as diffusion with drift [10] uses the blood as medium to carry the information molecules which limits their scope of communication inside the circulatory system. They cannot be used if communicating ends do not have direct access to the blood stream.

The human body spans larger distances and exhibits many factors other than diffusion and drift that influence MC, such as ADME. A wider and more comprehensive method of MC is needed to enable communication between any points across the body systems. Therefore, we propose a new Molecular Signaling Method in the body, denoted MoSiMe. The method takes advantage of how the ADME bodily processes affects drugs, referred to as Pharmacokinetics (PK) in pharmacology [12], to enable MC between any points of the human body regardless of their distance even if they are in different parts of the body and separated by tissues and fluids. For example, it can be used to convey information from a transmitter in the gut and a receiver in the heart.

We further describe the architecture, design, and different components of a system of Molecular Communication in Human Body (MoCoBo) with its underlaying MoSiMe. The system is capable of reliably communicating information in the body using molecules. An analytical model for the body channel is derived and an experimental platform is built to prove the concept of using MC in the body, which also allows the testing of new MC components such transmitters, receivers, and modulation techniques. The experimental platform is implemented and verified against clinical trials, animal tests, and analytical models.

The paper contributions can be summarized as follows:

- We propose MoSiMe and use it to develop the MoCoBo system, which includes the system architecture and various active and passive MoCoBo transmitters and receivers. We show the functionality of MoCoBo analytically and experimentally. To the best of our knowledge, we are the first to introduce a fully functional molecular communication system in the human body, derive an analytical model that accurately characterizes it, and build a practical experimental platform to test it. The system and method can be characterized as a Linear Time Invariant (LTI) system and communication engineering tools are utilized to derive its response and predict its behavior. Our work provides a simple MC signaling method that avoids the challenges of using chemical signals such as in [13] which is primarily attributed to the utilization of two chemicals to perform MC. Complex chemical interactions complicates the characterization of the system and forces the use of complex detection algorithms that could be difficult to implement at receivers with limited processing capabilities. Furthermore, our work is not limited to the use of a single substance for realizing MC such as in [14], where the proposed system relies solely on magnetized nano particles and limits the selection choices of biosensors for detecting nano particles without providing an analytical model or verifying the method for use in the human body.
- We provide an analytical model for the MoCoBo system and method, by deriving the impulse response and channel model for the human body, which considers the effects of natural body processes on the channel. Unlike previous research efforts, we present a physical channel model for the human body that captures natural processes including absorption, distribution, elimination, and excretions. The presented model, which is adapted from Pharmacokinetics (PK), a branch of pharmacology that studies how drugs move within the body [12], also predicts molecule concentration changes at all parts of the body. While PK models simulate the concentration of drugs in the body over time, they are not adequate for designing, testing and analyzing a MC communication system. However, adapting PK models allows for modeling the entire body as an LTI system, thus enabling the derivation of the body impulse response and application of communication engineering tools to characterize the MC physical channel within the body. All routes of administration are modelled by intravenous and extravascular models. In particular, the extravascular parameters can be accommodated to any route to the body that does not have access to blood stream such as oral, inhaling, or muscular routes.
- We have created a MoCoBo experimental platform that mimics the natural process of the body to prove the concept of MC in human body. It can also be used for testing and improving future MoCoBo transmitters, receivers, channel codes, and modulation techniques. We have validated the experimental platform against clinical trials, animal tests, and the derived analytical models.

The rest of this paper is organized as follows. Section II reviews relevant literature. Section III provides an overview of pharmacokinetics and Section IV presents the body impulse response. Section V discusses the proposed MoCoBo system and method while the MoCoBo experimental platform is discussed in Section VI. The derived physical channel model and the experimental platform along with the MoCoBo system and method are validated in section VII. Summary, conclusions and future work are provided in Section VIII.

## II. LITERATURE REVIEW

### A. MC Channel Characterization

Many models have been developed in the literature to characterize MC physical channels and propagation. These models can be broadly classified, based on distance, into short, medium, and long range channel models [9]. Some of the short and medium range models are presented in [13-19]. Specifically, models for propagation via diffusion and flow assisted diffusion are presented in [10] and [15], while on chip propagation via active transport is discussed in [16] and [17], and bacteria assisted propagation is proposed in [11]. On the other hand, calcium signaling via gap junctions is modeled for different body tissues in [18], and MC channel models for different flow-based microfluidic channels were derived in [19].

While short and medium range channel models capture the dynamics of MC channels in the scale of a few millimeter or less, the human body spans larger distances and exhibits other factors that influence MC channels such as chemical reactions, absorption, distribution, metabolization, and excretion. Therefore, such models cannot be relied on for MC channel modeling the human body.

In [20], a long range MC channel model for the cardiovascular system is presented. The model estimates the blood velocity profile and the delivery rate of injected molecules (particulate drugs) at different locations in the blood vessel network using physiological parameters such as the blood pressure profile, cardiac input, and heartbeat stroke volume and rate profile. However, it does not incorporate the effect of body processes such as molecule elimination by the kidney or transformation by the liver. In addition, it does not describe what happens to molecules in other body tissues outside the cardiovascular network. Furthermore, it assumes molecules are input into the body intravenously and does not consider other routes of administration such as oral or muscular administration.

*B. MC Experimental Platforms*

Clinical trials or even animal tests are prohibitively expensive and difficult to access for engineering research of MC in body. So far, very few experimental platforms and testbeds have been built to demonstrate and test MC systems in gaseous or aqueous mediums. In gaseous mediums, the first experimental platform that realized manmade MC is presented in [21] and [22]. It employs MC to send text messages in the air over a few meters for the first time. In [23], the platform is used to compare the performance of MC and electromagnetic waves in infrastructure monitoring applications where it demonstrated that MC connections remained successful while electromagnetic waves connections failed in metallic air ducts. Based on the work of [21], a new platform that supports multiple input multiple output communication is developed in [24]. Similarly but in aqueous environments, in [13] and [14] two testbeds are built to test MC in vessels. In [13], a multi-chemical MC platform that uses acids and bases to modulate information in water is presented while magnetic nano particles are used as information carriers in [14]. However, While the testbeds in [13] and [14] consider testing parts of the body vessels, our experimental platform is built to mimic the human body environment and can provide prediction of MC status at different parts of the body and test different routes of MC singling into the human body. In addition, its predictions are validated against clinical trials, animal tests and analytical models.

## III. OVERVIEW OF PHARMACOKINETICS

Pharmacokinetics (PK) is a branch of pharmacology that studies how the body affects molecule movements and concentration of Administered Substances (ASs), such as drugs, food additives or cosmetics to the body over time. PK models aim to mathematically predict the concentration profile over time for an AS affected by the bodily processes impacting the AS concentration in the body.

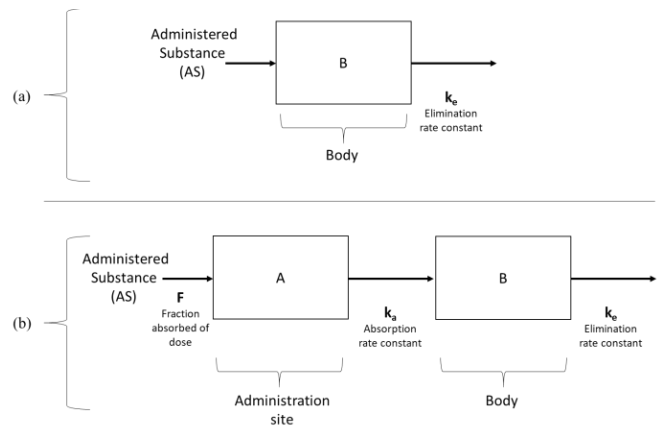

Fig. 1. One-compartment PK Model, (a) intravenous and (b) extravascular administration.

*A. Bodily Processes:*

The main bodily processes that affect ASs are absorption, distribution, metabolism and excretion (ADME). These processes determine the concentration of ASs in the blood and how it changes over time [25]. Once an AS is administered to the body, the absorption process moves the AS from the administration site to the circulatory system. There are many routes that an AS can take to enter the body such as oral, dermal, intravenous, or inhalation. The absorption rate depends on the chemical properties of the AS and the route of administration. From the route of administration, an AS diffuses passively through cells and membranes to reach the circulatory system which determines the extent of the absorption process. However, the absorption process is bypassed if an AS is administered intravenously as it directly enters the circulatory system. The blood then carries the AS and distributes it throughout the body. Once an AS reaches the blood, the metabolism and excretion processes start to eliminate the AS from the body and, therefore, the two processes are sometimes combined in a single process denoted elimination. The elimination process continues till the AS is completely removed from the body or the natural level of the AS in the blood before administration takes place is retained to insure the body's stability or homeostasis. The metabolization process denotes the biotransformation of chemical structure of an AS into other substances which mainly takes place in the liver. The excretion process filters ASs from the blood and is preformed mainly by the kidneys and partially in the liver.

*B. PK Modeling:*

PK models mathematically estimate the concentration levels of an AS in the body over time [25]. The model represents the body as a set of interconnected compartments. It assumes that the rate of transfer of an AS into compartments and the rate of elimination from compartments due to ADME processes follows linear or first-order kinetics. The number of compartments is decided based on the observation of the AS concentration profile over time which depends on the AS physical and chemical properties and how the ADME processes act on it. In this study, we make two assumptions about ASs used for MC signaling. First, ASs are selected so that their physical and chemical properties allows them to rapidly distribute in the body. Therefore, the distribution process will not affect ASs and, more importantly, the model is simplified to enable us to represent the entire body as a single homogenous compartment that follows the one compartment PK model. While physiologically real AS concentrations in organs or tissues will be different than the estimated levels by the one compartment model, the model estimation of the change in concentration will quantitively reflect changes (increasing or decreasing) in all tissues. Second, ASs are not bio transformed or metabolized, therefore, there will be no metabolization effect on our MC signal.

Fig. 1 (a) and (b) show the one compartment PK model for intravenous and extravascular administrations respectively. ASs are assumed to be eliminated from the body in the first-order process with elimination constant $k_e$. The rate of dose elimination is proportional to the quantity of the dose remaining in the body. Depending on the rate and extent of absorption, the amount of AS in the body can be estimated based on the route of administration to the body.

If an AS is administered in a place where it can reach the blood quickly, such as intravenous injection as shown in Fig 1 (a), the absorption process will have no effect on the AS, and therefore, the AS concentration would only be subject to the elimination process. Using Fig. 1 (a), the rate of change in the amount of AS in the body would be given by

$$\frac{dB}{dt} = -k_e B + IS \quad (1)$$

where IS is the input signal, B is the amount of the AS, and $k_e$ is the first order elimination rate constant for the AS. The negative sign indicates decrease in the number of particles of AS. Solving (1) as in [12] yields the equation that predicts concentration over time:

$$C_B = \frac{B_0}{V} e^{-k_e t} \quad (2)$$

where $B_0$ is the initial AS concentration at t = 0 and $V$ is the apparent volume of distribution.

On the other hand, if an AS is given extravascular as shown in Fig.1 (b), (e.g. orally), it would reach the blood slowly and the absorption process will control how fast it goes into the body by a first-order absorption constant $k_a$ that is specific to the site of administration. For example, $k_a$ for oral administration will be different from $k_a$ for dermal administration. As a result, the AS amount in the body is controlled by the absorption and elimination processes and the rate of change of the AS in the administration and body compartment is given by

$$\frac{dA}{dt} = -k_a A + F \, IS \quad (3)$$

$$\frac{dB}{dt} = k_a A - k_e B \quad (4)$$

where A is the amount of the AS at the administration site, $F$ is the absorbable fraction of $A_0$, and $k_a$ is the first-order elimination rate constant for the AS from the administration site. Solving (3) and (4) as in [12], we get

$$C_B = \frac{k_a F \, A_0}{V \, (k_a - k_e)} (e^{-k_e} - e^{-k_a t}) \quad (5)$$

where $A_0$ is total amount of AS administered at t = 0; and $V$ is the apparent volume of distribution.

The values for $k_e, k_a, F,$ and $V$ are estimated by obtaining the AS concentration profile over time for the patient and then using the method of residuals or the least squares curve fitting techniques [12].

IV. IMPULSE RESPONSE OF BODY

The PK model shown in Fig. 1 can be represented as an LTI system that consists of two internal LTI systems: the administration site compartment and the body compartment. The input to the system is a molecular signal and the output is the desired signal that conveys information. We are interested in finding the impulse response for two cases based on the location of the input signal introduction. The first case is intravenous administration when an input is directly injected into the body compartment, i.e., Fig.1 (a). The second case is extravascular administration when the input signal enters via the administration compartment, i.e., Fig.1 (b). Using the rate equations for each compartment, the impulse response of both cases are derived in this section.

This system is a continuous LTI system for which its input and output satisfy a linear constant-coefficient differential equation of the form

$$\sum_{k=0}^{M} a_k \frac{d^k y(t)}{dt} = \sum_{k=0}^{N} b_k \frac{d^k x(t)}{dt}. \quad (6)$$

Thus, the frequency response for each compartment can be obtained using the following equation from [26]:

$$H(j\omega) = \frac{Y(j\omega)}{X(j\omega)} = \frac{\sum_{k=0}^{M} b_k (j\omega)^k}{\sum_{k=0}^{M} a_k (j\omega)^k} \quad (7)$$

Considering the more general extravascular case, Fig. 1 (b), and using (3) at the administration site compartment, the frequency response is

$$H_{\text{Admin}}(j\omega) = \frac{F}{(k_a + j\omega)}. \quad (8)$$

Similarly, using (4), the frequency response of the body compartment is

$$H_{\text{Body}}(j\omega) = \frac{k_a}{(k_e + j\omega)}. \quad (9)$$

Therefore, the frequency response for the extravascular administration case is

$$H_e(j\omega) = H_{\text{Admin}}(j\omega) \, H_{\text{Body}}(j\omega)$$
$$= \frac{F}{(k_a + j\omega)} \frac{k_a}{(k_e + j\omega)}, \quad (10)$$

and the time domain impulse response for the total amount of the dose is

$$h_e(t) = \frac{F \, k_a}{(k_a - k_e)} (e^{-k_e t} - e^{-k_a t}). \quad (11)$$

Dividing by the volume we obtain the impulse response of the system for the concentration

$$h_{ec}(t) = \frac{F\, k_a}{V\, (k_a - k_e)} (e^{-k_e t} - e^{-k_a t}). \quad (12)$$

Similarly, the frequency response intravenous administration case is

$$H_i(j\omega) = H_{Body}(j\omega) = \frac{1}{(k_e + j\omega)}, \quad (13)$$

and the time domain impulse response for the total dose amount is

$$h_i(t) = e^{-k_e t}, \quad (14)$$

and dividing by the volume we obtain the impulse response of the system for the concentration

$$h_{ic}(t) = \frac{1}{V} e^{-k_e t}. \quad (15)$$

## V. PROPOSED MOSIME AND MOCOBO SYSTEM

Like any traditional communication system, the proposed MoCoBo system consists of three main components: a transmitter, a channel, and a receiver as shown in Fig. 2. However, it differs in using the body as the communication channel and employing a new MoSiMe for information signaling, which requires changes to the design of the traditional communication system components. In the following subsections, the channel and MoSiMe will be presented. Then, we will characterize the properties, design requirements, and modifications to the transmitter and the receiver to support the new signaling method.

### A. The Channel

The communication channel is the human body. Molecules travel from the transmitter's side to the receiver's side passively by drift and diffusion in the body liquids. The molecules propagate from site of release via diffusion till they arrive to the blood stream. Then, the blood carries and distributes them all over the body. Molecules diffuse slowly from the blood to different organs and tissues via means of diffusion based on concentration gradient. The body also starts to eliminate molecules from blood via elimination organs such as the liver, kidney and sweat glands which results in reducing molecule concentration in the body. Therefore, the concentration in the blood becomes less than other tissues and results in reverse diffusion of molecules from tissues to blood. The process of maintaining fixed levels of a specific concentration is called homeostats. The process continues till the normal level of molecular concentration is reached within the body.

### B. MoSiMe In the Body

The proposed system makes use of MoSiMe to send information through the body. MoSiMe exploits the effects of ADME processes on the AS to create a wave. Fig. 3 shows a typical wave generated in the body by this method where the concentration of AS molecules in the body represents the signal amplitude versus time. The wave has two phases: the rising phase and falling phase. The rising phase starts after injecting AS molecules into the body as the concentration starts to

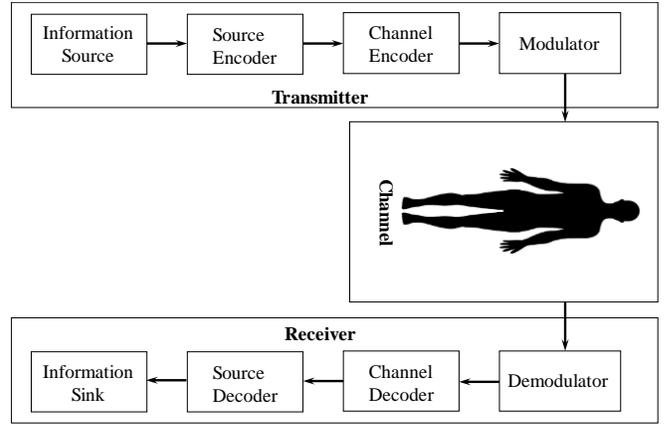
Fig. 2. System-level model for in-body molecular communication

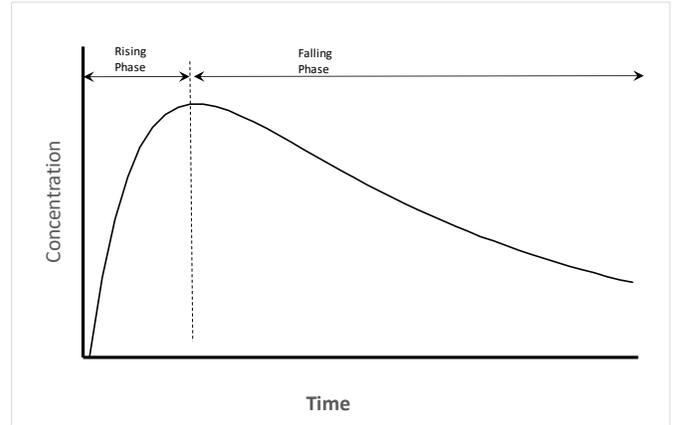
Fig. 3. Administered Substance (AS) concentration profile over time

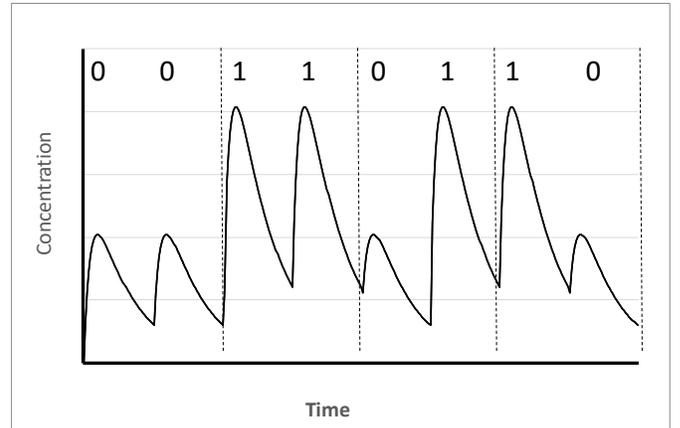
Fig. 4. Molecular signal

increase with a rate proportional to how fast the signaling molecules get absorbed by the body and reach the circulatory system. As the signaling molecules start entering the blood system, the body starts eliminating them out of the body by elimination organs such as the liver, kidneys or sweat glands or break them down into other compounds. If the injection of AS molecules stops, the elimination process starts the falling phase and leads to a decline in the signaling molecules' concentration. By exploiting the bodily processes of absorption and elimination, a molecular wave can be generated by varying the number of molecules released at the transmitter with time which creates a signal that can modulate and carry information across the body as shown in Fig. 4.

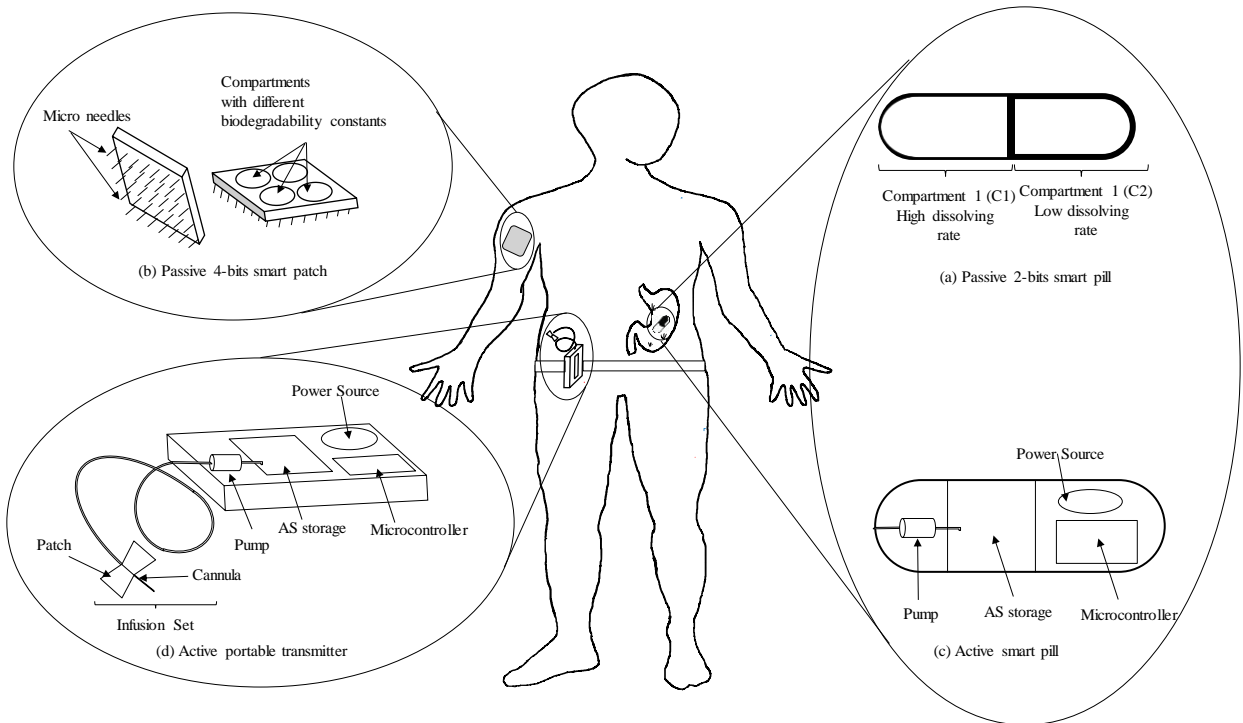

Fig. 5. Different MoCoBo transmitters; (a) Passive 4-bits smart patch, (b) passive 2-bits smart pill, (c) Active smart pill, (d) Active portable transmitter

*C. The MoCoBo Transmitter:*

From the logical design point of view, The MoCoBo transmitter components are similar to any traditional transmitter components. It has an information source, a source encoder, a channel encoder, and a modulator as shown in Fig. 2. The information source has a set of messages to be sent to the receiver. The source encoder takes a message, converts it to its binary representation and passes it to the channel encoder where extra bits are added for error detection and correction. Streams of bits are then modulated and sent into the channel.

The MoCoBo transmitter can be made of active components such as traditional transmitters or passive components which is an advantage of using MoSiMe for signaling.

An active transmitter consists of a microcontroller, power source, AS reservoir and a releasing mechanism. The microcontroller is powered by the power source and is programmed to perform all the functions of a traditional transmitter except that it should have a mechanism to control the AS amount and release time based on the information to be modulated. The release mechanism could be controlled using a pump that is powered to administer molecules and switched off to stop.

On the other hand, a passive transmitter could deliver a molecular signal without electrically powered parts utilizing physical and chemical properties of the body and the transmitter's components. For the transmitter to generate a molecular signal that modulates information, it needs to control the AS released amount and time of release. This could be achieved using multiple compartments made of biodegradable materials. The compartments would hold different concentrations of the AS inside biodegradable walls with different dissolution time. The molecular signal can then be modulated by varying concentrations inside compartments, with the biodegradable walls controlling the time of release.

Fig. 5 (a) shows an example of a passive 2-bits smart pill that can modulate two bits. The pill contains two compartments, denoted $C_1$ and $C_2$. Each compartment contains a solid or liquid form of an AS. The compartments are covered with membranes with different biodegradability constants and are made of different materials so that the dissolution time $T_1$ of $C_1$ is greater than the dissolution time $T_2$ of $C_2$. To modulate two bits of information, we need to have two levels of concentrations of the AS, denoted $L_0$ and $L_1$, where $L_0$ is half of $L_1$. By setting different combinations of $L_0$ and $L_1$ in the compartments, we can generate the molecular signal corresponding to 00, 01, 10, and 11 as shown in Fig. 4.

There are many routes a molecular signal can take into the body, such as oral, dermal, intravenous, or inhaling or any other route used by regular medications. The molecular signal route of administration determines the MoCoBo transmitter form. For example, in oral administration the transmitter components can be enclosed in pill shaped containers so that they can be taken orally and work from inside the body as shown in Fig. 5 (a) and (c). Alternatively, dermal route can be used where the transmitter should be designed to work from out of the body by attaching it to the skin such as transmitters in Fig 5 (b) and (d). A passive transmitter can be realized in the form of a patch that has compartments storing the signaling molecules and controlling their release time using biodegradable material. The molecules would be carried from the compartments and injected into the body via micro needles. Fig. 5 (b) shows a passive 4-bits smart patch.

An active transmitter can be placed inside a small box with an infusion set that has a cannula is shown in Fig 5 (d). The cannula is inserted through the skin till it reaches a fatty tissue and a patch is used to hold the cannula in place while delivering the AS to the body.

There are advantages and disadvantages for passive and active transmitters. While passive transmitters are ideal for sending one-time short and pre-encoded messages, active transmitters have the flexibility to send many messages with longer lengths. Furthermore, active transmitters can support full duplex when equipped with a receiver module. However, active transmitters have higher costs and require a power source.

*D. The MoCoBo Receiver:*

The MoCoBo receiver has the same logical components of a traditional receiver: demodulator, channel decoder, source decoder, and an information sink as shown in Fig. 2. The receiver can be an Implanted Electronic Medical Device (IEMD) that has a microcontroller, power source, and a bio sensor. The microcontroller detects signals received by the biosensor and estimates the channel symbols. It then runs the channel decoder to detect and correct errors in the transmitted stream of bits. After that, the source decoder predicts the message sent based on the estimated stream of bits.

The main difference between the physical components of the MoCoBo receiver and traditional receivers is the demodulation method. MoCoBo receivers must be equipped with a biosensor that is capable of detecting the presence/absence of AS or measure the quantity or concentration of ASs. Therefore, the selection of ASs will dictate the appropriate type of biosensor used. For example, a metal-oxide sensor, currently used in an IEMD for measuring signals generated by glucose (sugar), could be placed in the interstitial tissue to measure glucose concentration [27], [28].

## VI. MoCoBo Experimental Platform

We built an experimental MoCoBo platform to test the proposed MoCoBo system and method under constraints that mimic the human body environment. The platform can be used to test and verify new transmitters, receivers, or modulation techniques. The platform consists of two parts: the body modeling components and the communication ends components as shown in Fig. 6.

*A. Body Modeling Components*

The body modeling components consist of hardware and software parts that mimic the body physiological processes affecting a molecular signal. The current setup of the platform models two ADME processes which are the absorption and elimination that follow one-compartment PK model. It is also capable of testing different administration routes to the body such as oral and intravenous. The platform can be extended to model more complex PK models and ADME processes by adding more pumps and connections.

We first present the assumptions used to build the platform, which constraints the selection of the signaling molecules. Then, we derive the equations for calculating the required flow rates between compartments to generate ADME parameters such as absorption and elimination constants (i.e. $k_a$ and $k_e$). Finally, the platform setup and configurations are discussed.

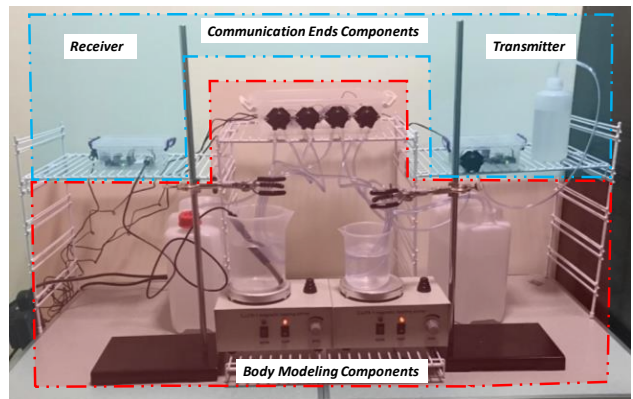

Fig. 6. MoCoBo experimental platform.

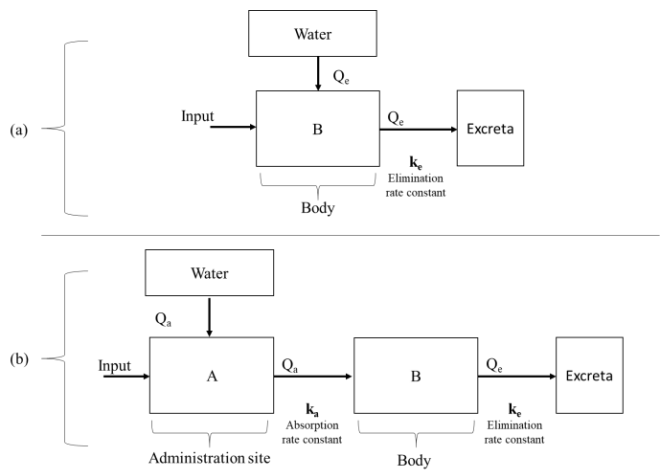

Fig. 7. Experimental platform logical diagram of the body components for (a) intravenous administration and (b) extravascular administration.

*1) Assumptions:*

In designing the experimental platform, the following assumptions are made:
- The signaling molecules get into the body and reach equilibrium rapidly.
- Mixing the solution in the compartment makes the signaling molecules homogenous all over the compartment instantaneously.
- The signaling molecules are not deformed or changed into other forms through metabolization or chemical reactions. Therefore, the rate of change in the concentration of the signaling molecules is directly proportional to its concentration in each compartment.

The first two assumptions are widely accepted in pharmaceutical sciences for modeling drugs using a one-compartment PK model [12],[29], which we have used as the basis for building this experimental platform. The third assumption is made to simplify the design of the experimental platform. However, the experimental platform can be extended to account for cases where the signaling molecules do not satisfy this assumption, by adding an output pump that accounts for loss of signaling molecules through metabolization or chemical reactions.

*2) Calculating Flow Rates*

Each AS has its absorption and elimination constants (i.e. $k_a$ and $k_e$) that reflects how fast the body absorbs or eliminates it.

Therefore, we need to find the flow rates (denoted by $Q_a$ and $Q_e$) and compartments' volumes (denoted by $V_a$ and $V_b$) that correspond to those constants. Next, the relationship between absorption and elimination constants, flow rates, and compartment volumes will be derived for the case of extravascular administration. The same approach can be used for the intravascular administration, which will lead to the same relationships and hence that is omitted for brevity.

From Fig. 7 (b), we derive the mass balance equations that determines the flow rates in terms of their corresponding constants. By the law of mass conservation [30], the mass balance equation for the administration compartment can be written as

$$\frac{dA}{dt} = m_w - m_a + m_I \quad (16)$$

where $A$ is the total mass accumulated in the administration compartment while $m_W$, $m_a$, and $m_I$ are the rates of mass flow from the water compartment into the administration, from administration to the body compartment, and from the input into the administration compartment, respectively. Similarly, the mass balance equations for the body compartment can be written as

$$\frac{dB}{dt} = m_a - m_e + m_w \quad (17)$$

where $B$ is the total mass accumulated in the body compartment while $m_a$, $m_e$, and $m_w$ are the rates of mass flow from the administration compartment into the body compartment, from the body compartment into the excreta compartment, and from the water compartment into the body compartment, respectively.

At the beginning of the experiment and at steady state, $m_I$ is zero. Noting that the flow of mass is

$$m = Q\,c \quad (18)$$

where $Q$ is the volumetric flow in mL/sec and $c$ is the concentration in particles per millions (ppm). Equations (16) and (17) can be rewritten as follows

$$\frac{dA}{dt} = Q_a\,c_w - Q_a\,c_a \quad (19)$$

$$\frac{dB}{dt} = Q_a\,c_a - Q_e\,c_b + Q_a\,c_w \quad (20)$$

Where $c_a$ and $c_b$ are the concentration in the body and administration compartments, and $Q_a$ and $Q_e$ are the volumetric follow rates from the administration to the body compartment, and from the body to the excreta compartment, respectively. To maintain constant volumes in the administration and body compartments, the volumetric flow rates from the water compartment to them is set to be equal to the outflows from the compartment. However, $c_w$ is zero because the flow coming from the water compartment carries no signaling molecules. Therefore, (19) and (20) reduce to

$$\frac{dA}{dt} = -Q_a\,c_a \quad (21)$$

$$\frac{dB}{dt} = Q_a\,c_a - Q_e\,c_b. \quad (22)$$

We also know that

$$A = c_a\,V_a \quad (23)$$
$$B = c_b\,V_b \quad (24)$$

where $V_a$, and $V_b$ are the concentration and volume in the body and administration compartments, respectively. Substituting (23) and (24) in (21) and (22), and comparing that outcome with (3) and (4), we can be obtain $Q_a$ and $Q_e$ as:

$$Q_a = k_a V_a \quad (25)$$

$$Q_e = k_e V_b. \quad (26)$$

*3) Setup*

The platform consists of a set of beakers, magnetic stirrers, Arduino controllers, peristaltic pumps and connecting tubes as shown in Fig. 6. While any liquid that represents the body fluids such as blood can be used in the platform, water is used as the carrying medium since its properties are very similar to body fluids, primarily as it constitutes more than 60% of body mass and it is the main element of its fluids. In addition, it is easier and cheaper at this early stage of testing to deal with water compared with using real plasma or blood, which are not as accessible and require special care and treatment. However, we remark that the platform can use other carrier media, after appropriate adjustments to the calculation of flow rates as presented in the next subsection.

Figs. 7 (a) and (b) illustrate the logical diagram of the body components for intravenous and extravascular administrations, respectively. They show how different parts are connected to each other and the flow rates between the compartments. In the case of intravenous administration, one beaker is used to represent the body compartment, while in the case of extravascular administration, two beakers are used to represent the administration and the body compartments. To maintain uniform distribution of the contents in the administration and body compartments, a magnetic stirrer is placed under each one of them. A large container filled with clear water is used to supply clean water to the administration and the body compartments. A large empty container is used to represent the excreta that collect waste removed from the body compartment.

In the intravenous administration, shown in Fig. 7 (a), the dose goes into the body compartment and hence, is only subject to the elimination process. The elimination process is modeled by clearing water from the body compartment to the excreta container while adding an equal amount to the body compartment from the clean water container to maintain a constant volume of water in the body compartment. To achieve this, two pumps with equal flow rates (i.e. $Q_e$ as shown in Fig. 7 (a)) are used, where one introduces clear water to the body compartment and the other draws water from it to the excreta container. The pumps's flow rate is controlled by connecting them to a Qunqi L298N motor drive and controlling their speeds via software installed in the Arduino controller.

In the extravascular administration, shown in Fig. 7 (b), the dose goes into the administration compartment and then moves to the body compartment. Therefore, it is subject to absorption and elimination processes. To model the two processes, three pumps with the same flow rate are used. The first moves clear water from the water source to the administration compartment, while the second pump moves solution from the administration compartment to the body compartment. The third pump is used to move the solution from the body compartment the excreta.

Using (25) and (26) for a given $k_a$ and $k_e$, the volumes of the administration and body compartments ($V_a$ and $V_b$) are found and set while making the flow rates $Q_a$ and $Q_e$ equal. Since we use low cost pumps it is easier to achieve the same effect of having a given absorption and elimination constants ($k_a$ and $k_e$) by fixing the flow rates ($Q_a$ and $Q_e$) to the same value, then, finding the volumes of compartments ($V_a$ and $V_b$) for the selected values for $k_a$, $k_e$, $Q_a$, and $Q_e$. The pumps' flow rate is again controlled using a Qunqi L298N motor drive controlled by software installed in the Arduino controller.

At the beginning of every experiment, clean water is added to the administration, body, and water compartments and a sample is taken as a blank and its Total Dissolved Solids (TDS) reading is used as a zero for all the subsequent measurements of concentrations in the testing environment.

### B. Communication End Components

An experimental MoCoBo transmitter and a MoCoBo receiver are built as follows to test MC in the experimental platform:

#### 1) The Transmitter:

The transmitter consists of a container, signaling liquid, a pump, an Arduino controller, and a power source as shown in Fig. 6. The container stores the signaling liquid and uses peristaltic pump geared with a 12 Volt DC motor. The pump is connected to a Qunqi L298N motor drive that is controlled by the PWM Arduino port to control its speed as needed. The Arduino is programmed to convert a given stream of binary bits into a series of ON-OFF pump cycles where ON runs the pump and the signaling liquid is injected by the transmitter while an OFF stop the pump. The duration of ON and OFF states depends on the encoding scheme.

Table salt is used as the source of generating signaling molecules because of its electrolyte properties when dissolved in water. The change of its concentration can be measured by the change of the conductance of the solution, which simplifies the task of signaling and detection. In addition, it is a common substance consumed in food and does not have harmful side effects or toxicity to human body if the amount consumed is within the recommended daily intakes. We remark that, even though table salt is used, the constructed experimental platform is generally applicable to emulate molecular communication with all signaling molecules by setting the correct flow rates, as demonstrated in our experiments presented in Sec. VI.A.3.

The main signaling molecules are Sodium ($Na^+$) and Chloride ($Cl^-$) ions plus other ions found in table salt such as iodine and calcium. Since it is not easy to control the release of precise amounts of solid form table salt, its liquid form is used by dissolving it in purified water.

To prepare the signaling liquid, one gram of Winsor table salt is dissolved in each liter of Aquafina purified water that has an average of four TDS. Then, the TDS of the solution is measured to determine the concentration of dissolved particles in the solution. The Atlas Scientific Conductivity (ASC) kit with a K 0.1 conductivity probe is used to determine the exact TDS of the formed liquid.

#### 2) The Receiver:

The receiver consists of a conductivity sensor, Arduino controller, and power supply. The conductivity sensor, made by ASC, is connected to the Arduino controller and can measure TDS in liquids with temperature compensation. The Arduino controller is programmed to log readings from the sensor and send them to MATLAB via a serial port.

## VII. RESULTS

### A. Validating the Analytical Model

To increase our confidence in the MoCoBo analytical model, its predictions for concentration levels over time are compared with clinical trial measurements and animal tests shown in Fig. 8 and Fig. 9 respectively. Fig. 8 shows the Chlorphenesin Carbamate plasma concentration levels over time in human plasma after oral administration of 3g [12]. Similarly, Fig. 9 shows measurements of the plasma concentrations of Vinpocetine after giving oral dose of 10mg/kg to a Wistar rat [31].

To validate the MoCoBo analytical model, the body is modeled as an LTI system with impulse response given by (14) where the values for *F*, *V*, $k_a$, and $k_e$ are obtained from [12] and [31]. The dose is modeled as an input signal in the form of an impulse with a magnitude equivalent to the dose weight. MATLAB is used to generate the body response by convoluting the input signal (dose) with the impulse response. The output of the convolution is the plasma concentration for the given drug under the given system model. The results show satisfactory match between the modeled and the observed measurements for both the human and rat data.

### B. Validating the Experimental Platform Against Human Tests

The predictions of the experimental platform are validated against the observed measurements of Chlorphenesin Carbamate oral dose given to a Human, which is shown also in Fig. 9.

Using (25) and (26), the body pump flow rates and compartment volumes are set as shown in Table I to generate the same of effect the absorption and elimination constants ($k_a$ and $k_e$) of Chlorphenesin Carbamate in a human. Then, a dose of 1 g of table salt is added to the administration compartment. The concentration of table salts are measured and scaled by a factor of 5.595 x $10^{-2}$ to mimic the effect of a 3g oral dose of Chlorphenesin Carbamate. The results show satisfactory match between the experimental platform and the observed measurements.

### C. Validating the Experimental Platform Against Animal Tests

The predictions of the experimental platform are validated against the observed measurements of Vinpocetine oral dose given to a Wistar rat, as shown in Fig. 9.

Using (25) and (26), the pump flow rates and compartment volumes are set as shown in Table II to generate the same effect of the absorption and elimination constants ($k_a$ and $k_e$) of Vinpocetine in a Wistar. Then, a dose of 522 mg of table salt is added to the administration compartment. The concentration of table salts are measured and scaled by a factor of $1.501 \times 10^{-3}$ to mimic the effect of a 10mg/kg oral dose of Vinpocetine. The results show satisfactory match between the experimental platform and the observed measurements.

*D. Validating the Experimental Platform Against the Analytical Model*

The analytical model is already validated against trial measurements and animal tests as shown in Fig. 8 and Fig. 9. Therefore, the experimental platform is validated against the analytical model predictions when testing the MoCoBo system.

For further testing, the analytical model is fed with the experimental platform parameters listed in Table III. The absorption and elimination constants ($k_a$ and $k_e$) are calculated using (25) and (26). The impulse response is then calculated for intravenous and extravascular administration using (11) and (14) respectively. To generate the impulse response for the experimental platform, an input signal should be created to resemble a delta signal that has infinite magnitude and zero width. To achieve this, a 130 mg dose of table salt is prepared. To reduce the width of the signal, the dose is put very quickly in a one-shot manner to generate the required impulse effect. The impulse signal is input into the body compartment to generate the impulse response for intravenous administration. The same procedure is repeated to generate the impulse response for the extravascular administration by releasing the impulse signal in the administration compartment.

The analytical and experimental normalized impulse response for the intravenous and extravascular administration are plotted in Fig. 10. which shows that the modeled and experimental impulse responses are matching.

*E. Testing MoCoBo Communication*

Our goal is to demonstrate that information can be sent using the proposed MoCoBo system and method. The main difference between the MoCoBo system and conventional communication systems lies in the use of molecular waves instead of electromagnetic waves while all other functionalities remain the same. Therefore, no testing is performed on common components such as the information sources and channel encoding in the transmitter and their counterparts in the receiver side. Instead, our focus is centered around testing the modulation, propagation, and demodulation components in the body using molecular communication where we aim to provide a proof of concept for the proposed system and method.

A predefined binary stream of bits (111-01010011) is modulated and demodulated using the proposed system and method. The binary stream consists of 11 bits where it starts by three ones to signal the start of communication, then 8 bits that represent the data part, which was selected to cover all possible combinations of two bits.

The modulation of bits is achieved by changing the concentration of the signaling molecules released in the body where a 1 is represented by releasing a dose and a 0 is represented by no dose. The dose representation for the input signal is plotted in Fig. 11, where the magnitude represents the amount of dose. Two scenarios are tested. In the first, the transmitter has direct access to the blood stream and hence the input signal is applied to the body compartment, thus mimicking an intravenous administration. In the second, the transmitter is assumed to be placed in the stomach or a muscle and therefore, the input signal is applied to the administration compartment to test extravascular administration.

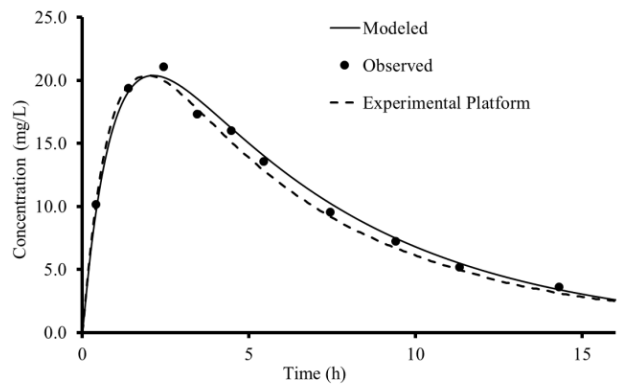

Fig. 8. Modeled vs. Observed vs. Experimental Platform Chlorphenesin Carbamate plasma concentration after giving oral dose of 3g to a Human.

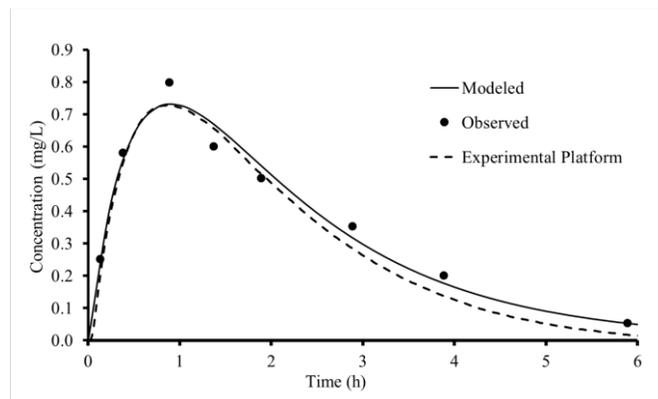

Fig. 9. Modeled vs. Observed vs. Experimental Platform Vinpocetine plasma concentration after giving oral dose of 10mg to a Wistar rat.

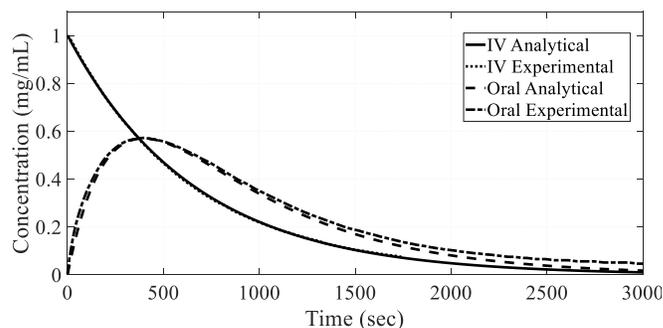

Fig. 10. The analytical and experimental normalized impulse response for the intravenous and extravascular administration

The modeled body response to the input signal is found by convoluting the prepared input signal with the impulse response. On the other hand, the experimental body response to the input signal is captured by measuring the concentration of the signaling molecules' levels at the receiver. The modeled and experimental body responses to the input signals are shown in Fig. 12 for the intravenous and extravascular administrations. We observe that the modeled response closely matches the experimental response despite some noise and spikes found in the experimental measurements. The spikes could be attributed to the mechanical noise generated by moving parts and switching the transmitter's pump on and off rapidly while modulating the input signal.

The input signal is recovered at the receiver's side by deconvoluting the body response with the impulse response of the body. Deconvolution is performed with the experimental impulse response and modeled impulse response, and their results are compared. As a reference for validating the recovered experimental input signal, the theoretical input signal is recovered by deconvoluting the modeled response with the modeled impulse response. The recovered signals are plotted in Fig. 13 for the case of intravenous administration and Fig. 14 for the case of extravascular administration. The recovered signal has sharp spikes in places where 1's are modulated and zero elsewhere, which resembles our choice of modulation. The magnitude of the recovered signal is less than the initial signal since in the current experimental setup, doses are introduced into the system by a pump with a constant flow rate which generates a pulse like signal rather than an impulse signal.

## VIII. SUMMARY, CONCLUSIONS AND FUTURE WORK

Recently, MC has been recognized as an enabling technology for nanonetworks [1], where MC is envisioned to enable nanorobots to achieve sophisticated and complex tasks in the human body for promising medical applications. Many MC methods between nanorobots in the human body have been proposed and modeled in the literature. Here, we propose a new method and system for Molecular Communication in the Body, denoted MoSiMe and MoCoBo, respectively. The method takes advantage of how the bodily ADME affects drugs, referred to as Pharmacokinetics (PK) in pharmacology, to enable MC in between any points of the human body regardless of their distance even if they are in different parts of the body and even if they are separated by tissues and fluids.

The MoCoBo system components and functionality are discussed in the context of traditional communication systems while explaining the differences and changes required due to the utilization of the new signaling method which we denote MoSiMe. MoSiMe employs ADME bodily processes such as absorption and elimination to realize signaling that conveys information between a transmitter and a receiver using molecules. Different designs for MoCoBo transmitters and receivers are proposed. Unlike traditional communication transmitters that must have active components, MoCoBo allows for creating passive transmitters that makes use of the chemical and physical properties of the signaling molecules and propagation environment. While traditional transmitters must work with a power source, we proposed for the first time in telecommunication design the concept of passive transmitters.

In addition, we adapt a physical channel model for the human body from pharmacology, where kinetics of drug and its behavior after giving a dose to a patient are studied. Our model describes the body as an LTI system characterized by the body physiological PK parameters and the signaling molecules released into the body are input to the system. The output of the system is the body response, which is represented by the concertation levels in the body over time. The model results are validated against two different drugs given orally to a human and a rat which show satisfactory match.

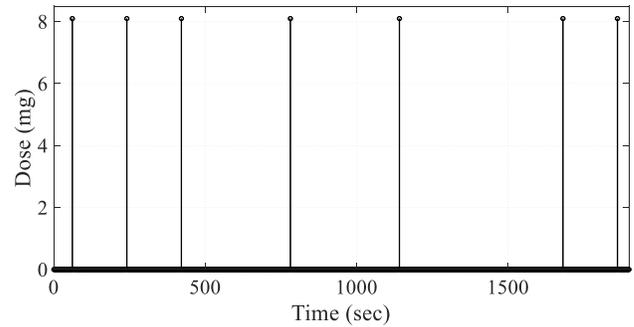

Fig. 11. Dose representation for the input signal 111-01010011.

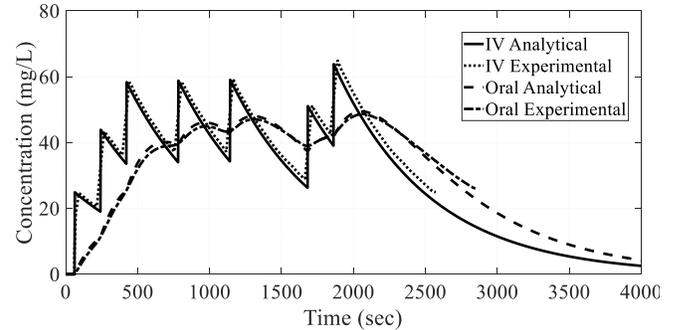

Fig. 12. The modeled and experimental body response to the input signal for intravenous and extravascular administrations.

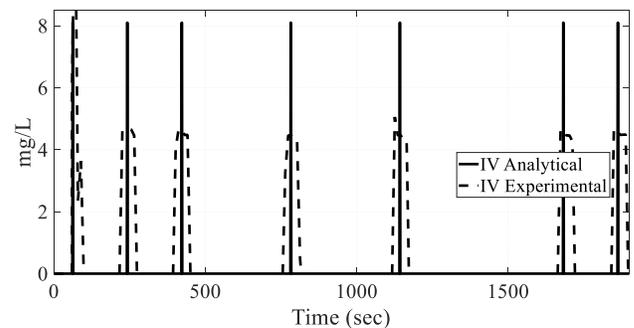

Fig. 13. The modeled and experimental recovered input signal for intravenous administrations.

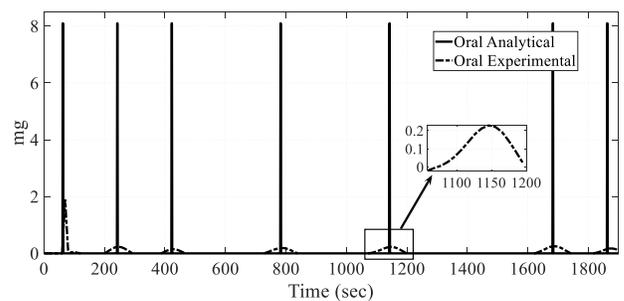

Fig. 14. The modeled and experimental recovered input signal extravascular administrations.

Conducting trials on human and animals is difficult for communication engineers as they lack the expertise and equipment in this field, which requires special facilities and procedures to perform. Therefore, we built an experimental platform to test the MoCoBo system and method. The platform consists of two components: one that models the body ADME processes and physical channel of the body while the other represents the communication ends. The platform is validated by comparing its response to clinical trials, animal tests and the analytical response of the body. One compartment PK model is tested using the platform and the results are matching the results predicted by the analytical model. Our results further show that messages can be sent between two devices using the proposed system. Two scenarios for transmitter placement are experimented based on the ability to access blood directly or indirectly, which we refer to as intravenous and extravascular administration. The first assumes that the transmitter has a release mechanism such as a needle connected directly to blood flow and can release the signaling molecules to it. The second assumes that the transmitter releases the signaling molecules in a part of the body that does not have direct access to the blood such as small intestines or muscles. The platform can be easily extended to support more complicated PK models with more compartments. It can also be used to test different modulation schemes, new MoCoBo transmitters, and new MoCoBo receivers.

TABLE I
EXPERIMENTAL PLATFORM PARAMETERS FOR PREDICTING CHLORPHENESIN CARBAMATE PLASMA CONCENTRATION IN HUMAN

| Symbol | Quantity | Value |
|---|---|---|
| $Q_a$ | Absorption flow rate | $1.025 \times 10^{-1}$ mL/s |
| $Q_e$ | Elimination flow rate | $1.025 \times 10^{-1}$ mL/s |
| $V_a$ | Volume of administration compartment | 355 mL |
| $V_b$ | Volume of body compartment | $2.292 \times 10^3$ mL |
| $k_a$ | Absorption constant | $2.89 \times 10^{-4}$ s$^{-1}$ |
| $k_e$ | Elimination constant | $4.47 \times 10^{-5}$ s$^{-1}$ |

TABLE II
EXPERIMENTAL PLATFORM PARAMETERS FOR PREDICTING VINPOCETINE CONCENTRATION IN WISTAR RAT

| Symbol | Quantity | Value |
|---|---|---|
| $Q_a$ | Absorption flow rate | $1.025 \times 10^{-1}$ mL/s |
| $Q_e$ | Elimination flow rate | $1.025 \times 10^{-1}$ mL/s |
| $V_a$ | Volume of administration compartment | 605 mL |
| $V_b$ | Volume of body compartment | 202 mL |
| $k_a$ | Absorption constant | $1.69 \times 10^{-4}$ s$^{-1}$ |
| $k_e$ | Elimination constant | $5.08 \times 10^{-4}$ s$^{-1}$ |

TABLE III
EXPERIMENTAL PLATFORM AND ANALYTICAL PARAMETERS FOR TESTING MOCOBO SYSTEM

| Symbol | Quantity | Value |
|---|---|---|
| $Q_a$ | Absorption flow rate | $9.8 \times 10^{-1}$ mL/s |
| $Q_e$ | Elimination flow rate | $9.8 \times 10^{-1}$ mL/s |
| $V_a$ | Volume of administration compartment | 650 mL |
| $V_b$ | Volume of body compartment | 300 mL |
| $k_a$ | Absorption constant | $3.27 \times 10^{-3}$ s$^{-1}$ |
| $k_e$ | Elimination constant | $1.51 \times 10^{-3}$ s$^{-1}$ |

There are many open issues that we propose as future work. The main issues can be divided into four categories. The first is related to the safety of using MC in the human body. Not any molecules will be suitable for signaling as some substances, when exceeding certain levels, might cause poisoning while others might interfere with natural body signaling processes. The second is related to data rates requirements of MC since it has very low data rates, which might require developing new communication protocols suitable for the available bandwidth. The third is related to transmitter design. In MC, passive transmitters can be created, which is not possible with electromagnetic communications. In addition, the placement of a transmitter in the body opens many issues and challenges. Moreover, a transmitter might have a limited amount of signaling molecules, which calls for efficient modulation techniques. The fourth category is related to the receiver design. The receiver must be equipped with a biosensor that matches the signaling molecules used by the transmitter, which opens the door for many choices and options.

## IX. REFERENCES

[1] L. Parcerisa Giné and I. F. Akyildiz, "Molecular communication options for long range nanonetworks," *Comput. Networks*, vol. 53, no. 16, pp. 2753–2766, Nov. 2009.
[2] J. T. Hancock, *Cell Signalling*. 2017.
[3] M. B. Miller and B. L. Bassler, "Quorum Sensing in Bacteria," *Annu. Rev. Microbiol.*, vol. 55, no. 1, pp. 165–199, Oct. 2001.
[4] B. Atakan, O. Akan, and S. Balasubramaniam, "Body area nanonetworks with molecular communications in nanomedicine," *IEEE Commun. Mag.*, vol. 50, no. 1, pp. 28–34, Jan. 2012.
[5] L. Felicetti, M. Femminella, G. Reali, and P. Liò, "Applications of molecular communications to medicine: A survey," *Nano Commun. Netw.*, vol. 7, pp. 27–45, Mar. 2016.
[6] S. J. Park *et al.*, "New paradigm for tumor theranostic methodology using bacteria-based microrobot," *Sci. Rep.*, vol. 3, no. 1, p. 3394, Dec. 2013.
[7] L. Felicetti, M. Femminella, G. Reali, and P. Liò, "A Molecular Communication System in Blood Vessels for Tumor Detection," in *Proc. of ACM The First Annual International Conference on Nanoscale Computing and Communication - NANOCOM' 14*, 2007, pp. 1–9.
[8] A. Guney, B. Atakan, and O. B. Akan, "Mobile Ad Hoc Nanonetworks with Collision-Based Molecular Communication," *IEEE Trans. Mob. Comput.*, vol. 11, no. 3, pp. 353–366, Mar. 2012.
[9] M. Gregori and I. Akyildiz, "A new nanonetwork architecture using flagellated bacteria and catalytic nanomotors," *IEEE J. Sel. Areas Commun.*, vol. 28, no. 4, pp. 612–619, May 2010.
[10] M. Pierobon and I. F. Akyildiz, "A physical end-to-end model for molecular communication in nanonetworks," *IEEE J. Sel. Areas Commun.*, vol. 28, no. 4, pp. 602–611, 2010.
[11] M. Gregori, I. Llatser, A. Cabellos-Aparicio, and E. Alarcón, "Physical channel characterization for medium-range nanonetworks using flagellated bacteria," *Comput. Networks*, vol. 55, no. 3, pp. 779–791, Feb. 2011.
[12] M. Gibaldi and D. Perrier, *Pharmacokinetics*. 1975.
[13] N. Farsad, D. Pan, and A. Goldsmith, "A Novel Experimental Platform for In-Vessel Multi-Chemical Molecular Communications," in *Proc. IEEE Global Communications Conference (GLOBECOM)*, 2017, pp. 1–6.
[14] H. Unterweger *et al.*, "Experimental Molecular Communication Testbed Based on Magnetic Nanoparticles in Duct Flow," in *Proc. IEEE International Workshop on Signal Processing Advances in Wireless Communications (SPAWC)*, 2018, vol. 2018-June, pp. 1–5.
[15] K. V. Srinivas, A. W. Eckford, and R. S. Adve, "Molecular communication in fluid media: The additive inverse gaussian noise channel," *IEEE Trans. Inf. Theory*, vol. 58, no. 7, pp. 4678–4692, 2012.
[16] S. Hiyama, Y. Moritani, R. Gojo, S. Takeuchi, and K. Sutoh, "Biomolecular-motor-based autonomous delivery of lipid vesicles as nano- or microscale reactors on a chip," *Lab Chip*, vol. 10, no. 20, pp. 2741–2748, 2010.
[17] A. Enomoto, M. J. Moore, T. Suda, and K. Oiwa, "Design of self-organizing microtubule networks for molecular communication," *Nano Commun. Netw.*, vol. 2, no. 1, pp. 16–24, 2011.
[18] M. T. Barros, S. Balasubramaniam, B. Jennings, and Y. Koucheryavy, "Transmission Protocols for Calcium-Signaling-Based Molecular Communications in Deformable Cellular Tissue," *IEEE Trans. Nanotechnol.*, vol. 13, no. 4, pp. 779–788, Jul. 2014.
[19] A. O. Bicen and I. F. Akyildiz, "System-theoretic analysis and least-squares design of microfluidic channels for flow-induced molecular communication," *IEEE Trans. Signal Process.*, vol. 61, no. 20, pp.

[20] Y. Chahibi, M. Pierobon, and S. O. Song, "A Molecular communication model of nanoparticle-body interactions in Particulate Drug Delivery Systems," in *2013 Asilomar Conference on Signals, Systems and Computers*, 2013, vol. 60, no. 12, pp. 1051–1055.
[21] N. Farsad, W. Guo, and A. W. Eckford, "Tabletop Molecular Communication: Text Messages through Chemical Signals," *PLoS One*, vol. 8, no. 12, p. e82935, Dec. 2013.
[22] N. Farsad, W. Guo, and A. Eckford, "Molecular communication link," in *2014 IEEE Conference on Computer Communications Workshops (INFOCOM WKSHPS)*, 2014, pp. 107–108.
[23] Song Qiu, W. Guo, S. Wang, N. Farsad, and A. Eckford, "A molecular communication link for monitoring in confined environments," in *2014 IEEE International Conference on Communications Workshops (ICC)*, 2014, pp. 718–723.
[24] B. H. Koo, C. Lee, H. B. Yilmaz, N. Farsad, A. Eckford, and C. B. Chae, "Molecular MIMO: From Theory to Prototype," *IEEE J. Sel. Areas Commun.*, vol. 34, no. 3, pp. 600–614, 2016.
[25] S. S. Jambhekar and P. J. Breen, *Basic Pharmacokinetics*. London: Pharmaceutical Press, 2009.
[26] A. V Oppenheim, A. S. Willsky, and S. H. Nawab, *Signals & Systems*, 2nd ed. Prentice Hall, 1996.
[27] M. Birkholz, P. Glogener, F. Glös, T. Basmer, and L. Theuer, "Continuously operating biosensor and its integration into a hermetically sealed medical implant," *Micromachines*, vol. 7, no. 10, pp. 1–8, 2016.
[28] D. Fitzpatrick, "Glucose Biosensors," in *Implantable Electronic Medical Devices*, Elsevier, 2015, pp. 37–51.
[29] A. B. C. Y. L. Shargel, *Applied Biopharmaceutics & Pharmacokinetics*, 7th ed. McGraw-Hill Education, 2016.
[30] L. Schmidt, Ed., *The Engineering Of Chemical Reactions*. OXFORD UNIVERSITY PRESS, 1998.
[31] R. S. Kadam, D. W. A. Bourne, and U. B. Kompella, "Nano-advantage in enhanced drug delivery with biodegradable nanoparticles: Contribution of reduced clearance," *Drug Metab. Dispos.*, vol. 40, no. 7, pp. 1380–1388, 2012.